\documentclass[
 reprint,https://v2.overleaf.com/project/5ba006002181f90dc7afddb8, superscriptaddress
]{revtex4-2}

\usepackage{amsmath}
\usepackage{amssymb}
\usepackage{graphicx} 
\usepackage{dcolumn}  
\usepackage{bm}       
\usepackage[caption=false]{subfig}
\usepackage{hyperref}
\usepackage{float}
\hypersetup{colorlinks=true, pdfstartview=FitV, linkcolor=blue, citecolor=black, plainpages=false, pdfpagelabels=true, urlcolor=blue}
\usepackage[all]{hypcap}
\usepackage[normalem]{ulem}

\begin{document}

\title{Effects of lattice dilution on the non-equilibrium phase transition in the stochastic  Susceptible-Infectious-Recovered model}

\author{Ruslan I. Mukhamadiarov} \email{mruslani@pks.mpg.de}
\affiliation{Max Planck Institute for the Physics of Complex Systems, Nöthnitzer Strasse 38, Dresden, D-01138, Germany}
\author{Uwe C. T\"auber} \email{tauber@vt.edu}
\affiliation{Department of Physics \& Center for Soft Matter and Biological Physics (MC 0435), Robeson Hall, 850 West Campus Drive, Virginia Tech, Blacksburg, VA 24061, USA}
\affiliation{Faculty of Health Sciences, Virginia Tech, Blacksburg, VA 24061, USA}

\date{\today}

\begin{abstract} 
We investigate how site dilution, as would be introduced by immunization, affects the properties of the active-to-absorbing non-equilibrium phase transition in the paradigmatic Susceptible-Infectious-Recovered (SIR) model on regular cubic lattices. 
According to the Harris criterion, the critical behavior of the SIR model, which is governed by the universal scaling exponents of the dynamic isotropic percolation (DyIP) universality class, should remain unaltered after introducing impurities. 
However, when the SIR reactions are simulated for immobile agents on two- and three-dimensional lattices subject to quenched disorder, we observe a wide crossover region characterized by varying effective exponents. 
Only after a sufficient increase of the lattice sizes does it becomes clear that the SIR system must transition from that crossover regime before  the effective critical exponents asymptotically assume the expected DyIP values. 
We attribute the appearance of this exceedingly long crossover to a time lag in a complete recovery of small disconnected clusters of susceptible sites which are apt to be generated when the system is prepared with Poisson-distributed quenched disorder.
Finally, we demonstrate that this transient region becomes drastically diminished when we significantly increase the value of the recovery rate or enable diffusive agent mobility through short-range hopping. \end{abstract}

\maketitle

\section{\label{sec:level1} Introduction}

The effect of quenched disorder on continuous phase transitions has been extensively studied in thermal equilibrium.
However, it remains to be understood if and how the character of a phase transition might change when quenched disorder is introduced to systems situated far from equilibrium. 
Significant efforts have been invested to study the influence of impurities on the active to absorbing state phase transition in variants of the contact process whose asymptotic universal properties are governed by the directed percolation (DP) universality class \cite{RevModPhys.76.663, doi:10.1080/00018730050198152, PhysRevE.55.6253, PhysRevE.57.5060, PhysRevLett.90.100601, Vojta_2006, PhysRevE.86.051137, Wada_2017, Kovacs2020, PhysRevLett.112.075702}. 
It was demonstrated that introducing quenched disorder to the contact or simple epidemic process leads to the appearance of Griffiths singularities indicative of regions that are devoid of impurities and that dominate the long-time density decay in the entire system \cite{PhysRevLett.57.90, PhysRevE.54.R3090, PhysRevE.69.066140, PhysRevE.72.036126, PhysRevE.90.042132}.
Beyond the contact process and its variations, the effects of quenched disorder have not been thoroughly investigated in other non-equilibrium systems. 
In this work, we thus investigate the non-equilibrium phase transition, namely the extinction transition for the associated epidemic, in the paradigmatic Susceptible-Infectious-Recovered (SIR) compartmental model that is implemented and simulated on two- and three-dimensional cubic lattices with site dilution, e.g., as a result of an immunization campaign.
In the absence of disorder, the critical behavior of a general epidemic process with recovery such as the SIR model should on a regular lattice typically be characterized by the scaling exponents of dynamic isotropic percolation (DyIP) \cite{GRASSBERGER1983157, Janssen1985, Cardy_1985, PhysRevE.82.051921, de_Souza_2011, JANSSEN2005147,Henkel:2008, Tauber14}. 

For phase transitions in thermal equilibrium, the famed Harris criterion may be invoked to predict possible changes in critical behavior due to quenched external perturbations. 
The Harris criterion dictates that when the corresponding pure system's correlation length exponent $\nu$ fulfills the inequality $d \, \nu > 2$ in $d$ spatial dimensions (i.e., if the specific heat critical exponent $\alpha = 2 - d \, \nu < 0$), the effects of disorder vanish in the thermodynamic limit upon successive coarse-graining \cite{Harris_1974}. 
Various generalizations of the Harris criterion have since been devised \cite{PhysRevB.27.413, Kinzel1985, Alonso_2001, PhysRevE.93.032143}. 
However, for quenched disorder that is uncorrelated in space and does not evolve with time, one can show that the Harris criterion maintains its original form \cite{PhysRevE.93.032143}.

According to the Harris criterion, the effects of disorder should hence disappear in the SIR model in the large system size limit in all dimensions $d$, see Table~\ref{tab:table} below (from Ref.~\cite{Henkel:2008}). 
However, in our detailed Monte Carlo simulations for the spatially extended stochastic SIR system we observe an extended power law crossover region for which the values of the scale-dependent effective exponents differ from the asymptotic DyIP critical exponents. 
Only in the very long-time limit do these effective exponents assume the expected ``clean" DyIP values. 
The prevalence of this large crossover regime hence poses the problematic issue that it may not be recognized (in either simulations or experimental / observational data) as such if the spreading infection front encounters the system's boundaries before the asymptotic long-time limit with universal scaling has been reached.
In the latter scenario, the effective exponents could be mistaken for and misinterpreted as the asymptotic critical exponents, albeit apparently assuming non-universal numerical values.
We demonstrate this point here by showing that site-diluted SIR systems find their asymptotic critical state which is characterized by DyIP exponents only after a sufficient and substantial increase in the size of the lattice.

We attribute the origin of the large crossover region to a time lag in the complete recovery of small disconnected clusters of susceptible individuals, which are inevitably generated when Poisson-distributed quenched disorder is introduced. 
As the percolation threshold density is approached and the infection rate increased to retain the system in a critical state, the infection propagates quickly.
Moreover, when the infection front reaches the perimeter of a small percolation cluster, most of the cluster's population will be in the infectious state. 
Such an infectious cluster then becomes extinct due to Poisson point processes, following a time delay which is inversely proportional to the recovery rate. 
Only once the fraction of independent simulation runs for which the seeds were placed inside small disconnected clusters goes extinct, the system eventually reaches its asymptotic behavior.
We show that upon drastically increasing the value of the recovery rate or upon enabling nearest-neighbor hopping dynamics (in addition to the SIR reactions), these long crossovers disappear. 
This finding supports our hypothesis about the crossover origin in the static scenario, since increase in recovery rate and the agents' mobility in the system both reduce the size and the effect of the time lag.

We end our introductory remarks with a brief overview of the recent pertinent literature to provide relevant background:
The effects of quenched disorder on a non-equilibrium phase transition in the DP universality class were investigated by Vojta and Lee by simulating the contact process on a regular lattice \cite{PhysRevLett.96.035701}. 
The authors demonstrated that the subtle interplay between varying percolation thresholds and dynamical fluctuations leads to a different universal behavior from the pure system. 
A modified version of the SIR model, where self-propelled agents are moving in a continuous space in the presence of quenched disorder, was recently studied by Forg{\'a}cs et al., who showed that lowering the infection transmissibility enhances the effects of disorder \cite{https://doi.org/10.48550/arxiv.2203.13341}.
Other types of disorder and their effects on critical properties of epidemic models were studied as well: 
For example, Wada and Hoyos added correlated temporal disorder to the Susceptible-Exposed-Infectious (SEI) model, which changed the universality class from dynamic isotropic percolation to an exotic infinite-noise universality class associated with the contact process \cite{PhysRevE.103.012306}. 
The effects of agent diffusion on critical properties of infection spreading in simple epidemic processes have also been widely investigated, see, e.g., Refs.~\cite{PhysRevA.39.2214, VANWIJLAND1998179}. 
For this class of models wherein a diffusive mode is coupled to a critical DP field, depending on the relative strength of the diffusivities for healthy and sick individuals, respectively, four different universality classes could be identified \cite{PhysRevLett.128.078302}. 
Finally, in recent work by \'Odor, the effect of quenched disorder was considered in a lattice SIR model by assigning different infectious rates to 10\% of randomly selected lattice sites \cite{PhysRevE.103.062112}.

This paper is organized as follows: 
In the following section we introduce the stochastic SIR model and its spatial extension on a regular lattice. 
We also outline the details of our Monte Carlo simulation algorithm. 
In Sec.~\ref{sec:level3} we present extensive and detailed simulation results for the SIR model on diluted two-dimensional and three-dimensional lattices with static agents.
Subsequently we discuss the modifications caused by allowing nearest-neighbor hopping, i.e., diffusive transport, of the agents. 
We conclude in Sec.~\ref{sec:level4} with a summary and discussion of our main results.

\section{\label{sec:level2} Model Description}

The Susceptible-Infectious-Recovered (SIR) compartmental model is a paradigmatic example for an epidemic process with recovery. 
The entire population in this model is divided into three compartments: S -- susceptible, I -- infectious, and R -- recovered and immune. 
The dynamics of the system can be represented by irreversible transitions from the S to I and I to R states. 
This transfer of individuals from one compartment to another can be represented by a set of chemical reactions: 
The infection reaction $S + I \to I + I$ which occurs with rate $r$, and the recovery reaction $I \to R$ with rate $a$. 
Note that these stochastic processes leave the total population number $N = S + I + R$ fixed.
Within a mean-field rate equation approximation, wherein a mass action-type factorization is applied to the two-body correlations in the infection processes, these reactions are described by a set of coupled ordinary differential equations for the three species:
\begin{align*}
\begin{split}
    \frac{dS}{dt} &= - \frac{\beta I S}{N} \, , \\
    \frac{dI}{dt} &= \frac{\beta I S}{N} - \gamma I \, , \\
    \frac{dR}{dt} &= \gamma I \, , 
\end{split}
\end{align*}
where $\beta$ and $\gamma$ are the continuum mean-field infection and recovery rates and $N(t) = S(t) + I(t) + R(t) = N(0)$ denotes the conserved total number of individuals.
Initially, we set $R(0) = 0$ and $I(0) \ll S(0) \approx N(0)$. 
An important epidemiological parameter that is associated with these rates is the basic reproduction number 
$\mathcal{R}_0 = \beta / \gamma$, which serves as the indicator that predicts an epidemic outbreak with $(dI/dt)|_0 > 0$, namely when $\mathcal{R}_0 > 1$. 

While the above SIR mean-field equations usually yield fairly accurate predictions for the course of an epidemic for large and well-connected populations, they cannot adequately account for the spatio-temporal fluctuations that originate both from environmental variability and the intrinsic stochasticity of the reaction processes, which may drive the system to an absorbing epidemic extinction state at the early stages of the infection outbreak \cite{Tauber14, doi:10.1142/9781786347015_0001, Mukhamadiarov2021}.
One way to account for these fluctuations is to write down a mesoscopic Langevin equation with multiplicative noise for the local density of infected agents, and proceed with the analysis through e.g., a perturbative treatment \cite{T_uber_2012} and if needed, renormalization and scaling analysis \cite{Tauber14}. 
Another generally applicable approach is to consider the stochastic implementation of the SIR reactions by simulating the individual processes $S + I \to I + I$ and $I \to R$ on an adequate spatial setting, such as a regular lattice; or more realistically for applications to disease spreading in human populations, on a network with appropriate topology and connectivity \cite{RevModPhys.87.925, doi:10.1098/rsif.2005.0051, 10.1371/journal.pone.0063935, Mukhamadiarov2021, Mukhamadiarov_2021, PhysRevE.103.062112}. 

In this work, we consider infection spreading on regular square and cubic lattices, modeled through individual-based Monte Carlo simulations for the stochastic SIR reactions. 
Depending on the values of the infection and recovery rates, the system will be in either the spreading or contained phases which are separated by the non-equilibrium phase transition. 
In the spreading or active phase, the SIR dynamics on a lattice can be viewed as a propagation of the infectious front emanating from a single infected nucleus into the entire domain populated by susceptible individuals. 
While the recovery process and ensuing memory effects play a crucial role in the SIR, especially when one considers its critical behavior, in the spreading phase the infection front moves outwards from the seed with a constant velocity and the front dynamics can be characterized by the Fisher--Kolmogorov--Petrovsky--Piskunov (FKPP) equation. 
The FKPP equation describes a situation when the front invades a linearly unstable state \cite{VANSAARLOOS200329}, and has been used to study, e.g., the population dynamics of a single species that undergoes spontaneous death and birth processes \cite{PhysRevLett.87.238303, Barreales_2020}; simple epidemic processes such as the SIS model \cite{murray2001mathematical}; as well as cluster growth processes as occurring in the Eden model \cite{HERRMANN1986153}. 
In the non-spreading or inactive phase, the infected population quickly approaches extinction, leaving the major part of the population in the susceptible state, unaffected by the disease; infection outbreaks remain localized and contained.

At the critical point and for asymptotically long times, all three population densities $\{S(t), I(t), R(t)\}$, the disease survival probability $P_{s}(t)$, i.e., the probability that at some given time $t$ the system has at least one infectious individual left, and the mean-square displacement of the spreading disease from its origin $\mathcal{R}^2(t)$ are known to exhibit power law scaling behavior \cite{stauffer1994introduction, GRASSBERGER1983157, Cardy_1985, PhysRevE.82.051921, de_Souza_2011, Tauber14, PhysRevE.56.5101, Dammer_2004, RevModPhys.76.663, https://doi.org/10.48550/arxiv.2208.12038}:
\begin{equation}
\begin{split}
N - S(t)& \sim \, R(t) \sim t^{\theta_\text{R}} \, , \hspace{0.2cm} 
I(t) \sim t^{\theta} \, ,  
\\[2ex]
P_{s}(t) &\sim t^{-\delta} \, , \hspace{0.4cm} 
\mathcal{R}^2(t) \sim t^{2/z} \,
\end{split}
\label{crtexp}
\end{equation}
with critical scaling exponents $\delta$, $\theta$, $\theta_\text{R}$ and $z$, where $\theta_\text{R} = d_\text{f} / z - \delta$ and $z$ is the associated dynamical critical exponent that captures critical slowing-down via connecting the divergent correlation length $\xi \sim |\tau|^{-\nu}$ with the characteristic relaxation time $t_c \sim \xi^z \sim |\tau|^{-z \nu}$, where $\tau$ denotes the distance from the critical point. 
The scaling relation for the exponent $\theta_\text{R}$ that governs the number of susceptible and recovered individuals originates from the following argument: 
The total mass of the percolating cluster is related to its linear extension via the associated fractal dimension $d_\text{f}$ through $M \sim L^{d_\text{f}}$, while the cluster's linear extension increases with time as $L \sim t^{1/z}$. 
The growth in the number of recovered individuals can then be obtained by multiplying the size of the percolating cluster by the probability that the cluster continues to grow after some time $t$, i.e., the survival probability $P_{s}$, which together produces $R(t) \sim t^{d_\text{f} / z - \delta}$.
For the dynamic isotropic percolation universality class, these exponents assume the universal values that depend only on the system's dimensionality $d$ listed in Table~\ref{tab:table} for two and three dimensions (from Ref.~\cite{Henkel:2008}). 
\begin{table}[!t]
\centering
\begin{tabular}{lclclclc}
\hline
\hline 
\rule{0pt}{2.5ex} & \hspace{0.5cm} $\delta$ & \hspace{0.8cm} $\theta$ & \hspace{0.5cm} $\nu$ & \hspace{0.7cm} $d_\text{f}$ & \hspace{0.5cm} $z$ \\
\hline \vspace{1mm}
\rule{0pt}{2.5ex} d=2 & \hspace{0.6cm} 0.092 & \hspace{0.5cm} 0.586 & \hspace{0.5cm}4/3 & \hspace{0.5cm} 91/48 & \hspace{0.2cm} 1.1295 \\
\rule{0pt}{2.5ex} d=3 & \hspace{0.6cm} 0.346 & \hspace{0.5cm} 0.488 & \hspace{0.5cm} 0.875 & \hspace{0.5cm} 2.523 & \hspace{0.2cm} 1.375 \\
\hline
\hline
\end{tabular}
\caption{\label{tab:table} The asymptotic critical exponents of the dynamic isotropic percolation (DyIP) universality class in $d = 2$ and $d = 3$ dimensions. 
(From Ref.~\cite{Henkel:2008}, p.~211.)}
\end{table}

We employ Monte Carlo simulations with random sequential updates to model the SIR epidemic spreading on regular two- and three-dimensional cubic lattices with periodic boundary conditions. 
Starting with a single infectious (I) seed which is placed in the lattice center, while the rest of the population is set in the susceptible (S) state, we proceed with randomly selecting lattice sites and attempting to perform the SIR reactions. 
If a chosen lattice site is occupied by an infectious agent, we try infecting all its susceptible nearest neighbors separately by picking a random number in the interval $[0,1]$ and turning a susceptible individual into the I state if the resulting number is less than the prescribed infection probability $r$. 
After we have tried infecting all susceptible nearest neighbors, we next attempt the recovery reaction for the chosen infectious (I) individual by calling the random number generator again and putting the individual into a recovered (R) state if the number that came out is less that the fixed recovery probability $a$. 
For one complete Monte Carlo step, we repeat this procedure of random lattice site selection and reaction attempts $V = L^d$ times for the two- ($d=2$) and three-dimensional ($d=3$) systems.
To simulate the SIR dynamics on a diluted lattice, we prepare the system by filling the lattice with agents in a random manner, so that the overall agent number $N$ corresponds to the chosen density $\rho = N / V < 1$. 
When we thus introduce quenched site dilution disorder to the system by filling the lattice in a random manner, we restrict the occupation number of a single lattice site to $0$ (empty) or $1$ (filled). 
Finally, we also consider the scenario where agents are allowed to hop to the nearest-neighboring lattice sites with some probability $p$, if those sites are not occupied. 
This endows the agents with intrinsic diffusive mobility in addition to the disease spreading dynamics.
For all our simulation results presented here we set the recovery probability to a fixed value $a = 0.1$, but vary both the infection and hopping probabilities $r$ and $p$.

We point out that there are in principle two distinct finite-size effects to be considered in any numerical SIR model simulation that inevitably cause a departure from the critical power laws (\ref{crtexp}).
First, there are of course cutoffs induced by the finite lattice extension, which become manifest when the intrinsic correlation length reaches the system size, $\xi \sim L$.
These finite-size effects are effectively suppressed in initial-seed simulations, and become relevant only when $\mathcal{R}^2(t) \sim L^2$, or once $t \sim L^z$. 
At that point the spreading infection front reaches the system boundaries, resulting in the characteristic infection curve peaks with maximum $I_m$ for $I(t)$ for the SIR and related models.
Second, generally in the SIR model with fixed total agent number $N$, epidemic spreading is naturally limited by that finite population number and the recovery of infected individuals to the R state. 
Since $I_m \sim N$ constitutes a certain fraction of the totally available population, the critical growth law with exponent $\theta$ must terminate well before $t \sim N^{1/\theta}$.
Yet in our simulations, with restricted site occupations, we have $N \sim L^d$, and hence these two finite-size effects are no longer independent.
Indeed, since $z > d / \theta$ according to a general scaling relation \cite{Tauber14}, the latter ultimately sets the limiting constraint for seed simulations.

\section{\label{sec:level3} Results}

\begin{figure}[t]\label{fig:fig1}
    \centering
    \includegraphics[width=1.0\columnwidth]{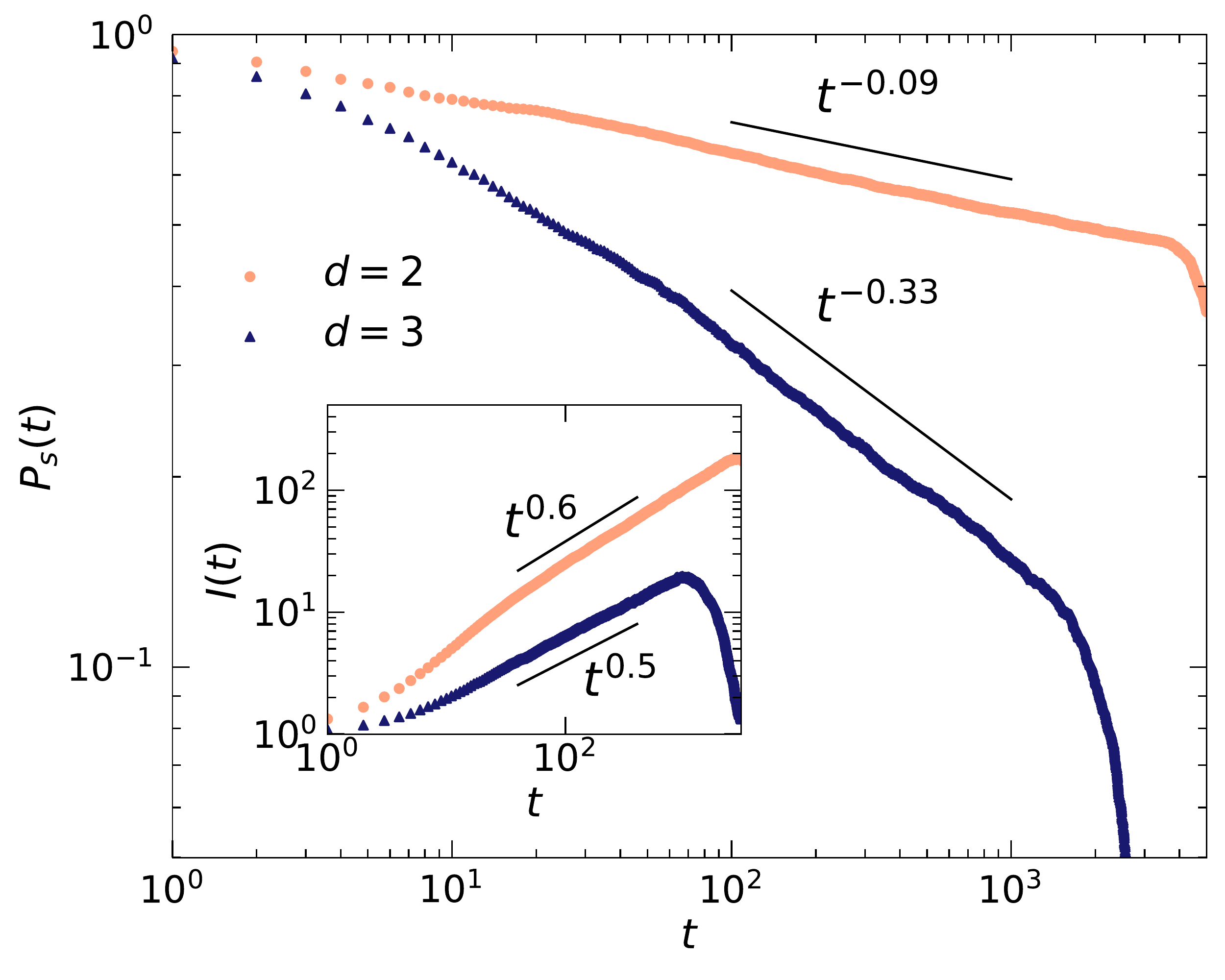}
    \caption{Time evolution of the infection survival probability $P_s(t)$ at the critical point (double-logarithmic plot). 
    The infection spread is being modeled by performing stochastic SIR reaction simulations on regular square and cubic lattices with periodic boundary conditions. 
    The inset depicts the time evolution of the total number of infected individuals $I(t)$ at the critical point. The lattice density is $\rho=1.0$, and the critical values of the infectious and recovery rates are $r_c = 0.1028$, $a_c = 0.1$ and $r_c = 0.0345$, $a_c = 0.1$ for $d=2$ and $d=3$ respectively. 
    The simulation system sizes are $500^2$ and $50^3$ for two- and three-dimensional lattices, respectively. Our best estimates for the exponents are $\delta = 0.09 \pm 0.01$, $\theta = 0.6\pm 0.02$ for $d = 2$ and $\delta = 0.33 \pm 0.02$, $\theta = 0.5 \pm 0.03$ for $d = 3$.
    The survival probability curves were obtained from averaging over $3000$ independent realizations; the infection growth curves in the inset were obtained from the very same simulation runs.}
\end{figure}

Before we proceed to investigating the effects of quenched disorder, we first confirm known results and calculate the values of the ``clean'' critical exponents that characterize the non-equilibrium active-to-absorbing phase transition of the stochastic SIR model on regular square and cubic lattices with periodic boundary conditions. 
To that end, we initialize our lattices with all sites occupied with single susceptible individuals $S$ (density $\rho = 1$), except for a single seed position which is set to be in the infectious state $I$. 
We then run the Monte Carlo simulations and measure the temporal evolution of the total number of individuals in each distinct state $S(t)$, $I(t)$, and $R(t)$, the survival probability $P_s(t)$, and the mean-square displacement $\mathcal{R}^2(t)$. 
We specifically select the survival probability to locate the critical value of the infection rate $r_c$ (keeping the recovery rate $a$ fixed).
Our choice is dictated by the fact that the $P_s(t)$ curve continues to decay with a power law even after the infectious curve hits its maximum $I_m$.
Indeed, the presence of the maximum in the $I(t)$ curve substantially limits the sensitivity in these data to determine the value of the critical infectious rate $r_c$, which is crucial for proper identification of the asymptotic critical exponents. 
Using the survival probability curve, we search for the value of the infectious rate $r_c$ that separates the graphs of $P_s$ for $r > r_c$ which asymptotically reach some non-zero values from the curves for $r < r_c$ that ultimately always tend to zero. 
To calculate the survival probability $P_s(t)$ from our simulation data, we simply keep track of the fraction of independent simulation runs (relative to the total number of submitted runs) that hit the absorbing state at the prescribed time $t$. 
Figure~\ref{fig:fig1} shows the results of our Monte Carlo simulations for completely filled two- and three-dimensional lattices. 
Our data, yielding $\delta = 0.09 \pm 0.01$, $\theta = 0.6 \pm 0.02$ for $d = 2$ and $\delta = 0.33 \pm 0.02$, $\theta = 0.5 \pm 0.03$ for $d = 3$, indicate that the asymptotic critical exponents take the standard DyIP values from  Table~\ref{tab:table} in both two and three dimensions, in full accord with previously reported results \cite{GRASSBERGER1983157}. 

\begin{figure}[t]
    \centering
    \subfloat{\includegraphics[width=\columnwidth, trim={0 0 0cm 0.4cm},clip]{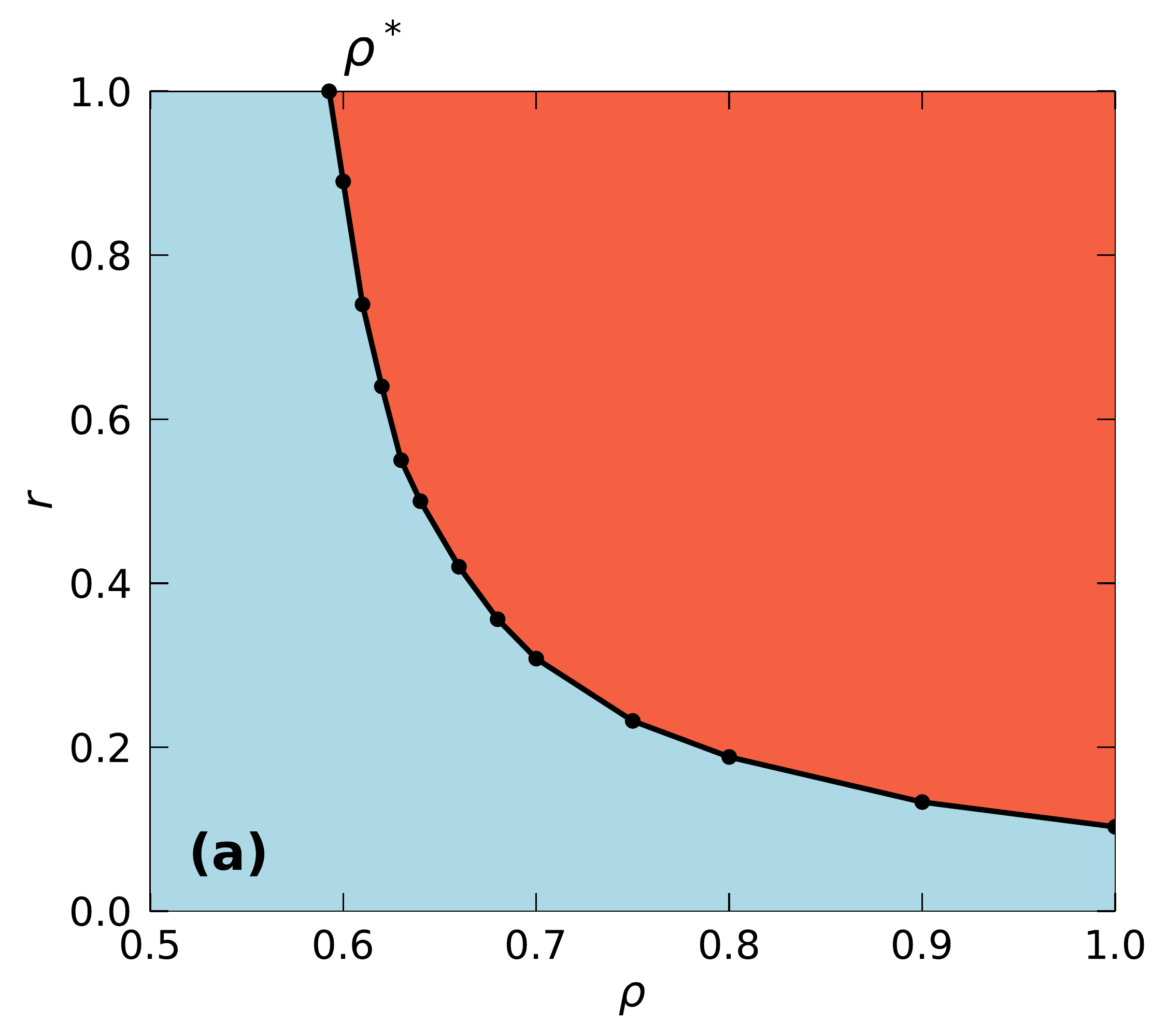}}
    \vfill
    \vspace{-0.5cm}
    \subfloat{\includegraphics[width=\columnwidth]{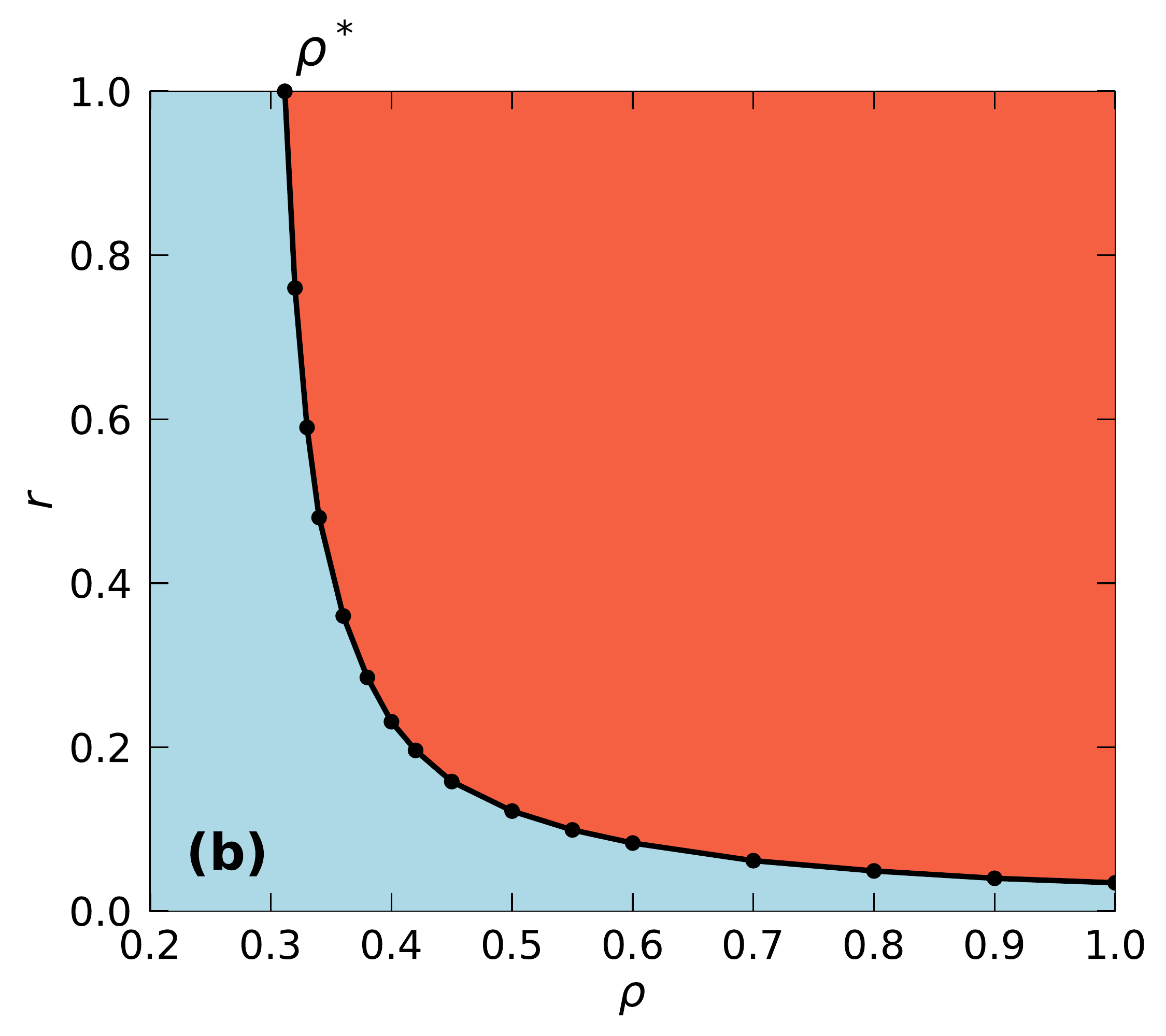}}
    \caption{Non-equilibrium phase diagrams for the stochastic SIR model with quenched site dilution disorder, simulated on (a) square ($d = 2$) and (b) cubic ($d = 3$) regular lattices.
    The active, spreading phase is colored in red (dark gray) and the absorbing, non-spreading phase in light blue (light gray). 
    The critical line was obtained from a linear interpolation between the critical values of the infectious rate $r_c(\rho)$ that were  located separately for each density value $\rho$. 
    The lower density limit $\rho^*$ for the active phase at $r_c = 1$ is set by the lattices' site percolation thresholds: $\rho^* \approx 0.5927$ and $\rho^* \approx 0.3116$, respectively, in two and three dimensions.} \label{fig:fig2}
\end{figure}
%
According to the Harris criterion in its original form, both sets of DyIP exponents for $d=2$ and $d=3$ satisfy the inequality $d \, \nu > 2$, where $\nu$ is the critical exponent associated with the correlation length $\xi$. 
Since the criterion is satisfied, one should expect that introducing quenched spatial disorder will not change the character of the non-equilibrium phase transition, and the effective exponents would asymptotically assume their ``clean'' values. 
To carefully investigate the effects of quenched site dilution randomness, we start reducing the overall density of agents starting from $\rho = 1.0$ until the site percolation threshold $\rho^*$ is reached. 
The latter restriction is necessary since for lattices with quenched site dilution, regardless of the rate $r > r_c$, the infection can only spread through connected clusters of susceptible individuals. 
Thus, the critical line $r_c(\rho)$ that we construct by linearly interpolating the critical values of the infection rates as function of the density $\rho$, must terminate at $\rho^*$. 
As shown in Fig.~\ref{fig:fig2}, this critical line separates two distinct non-equilibrium phases, the active state with disease spreading and the inactive, absorbing, non-spreading phase.

\begin{figure}[t]
    \centering
    \subfloat{\includegraphics[width=\columnwidth, trim={0cm 0 0.8cm 0.8cm},clip]{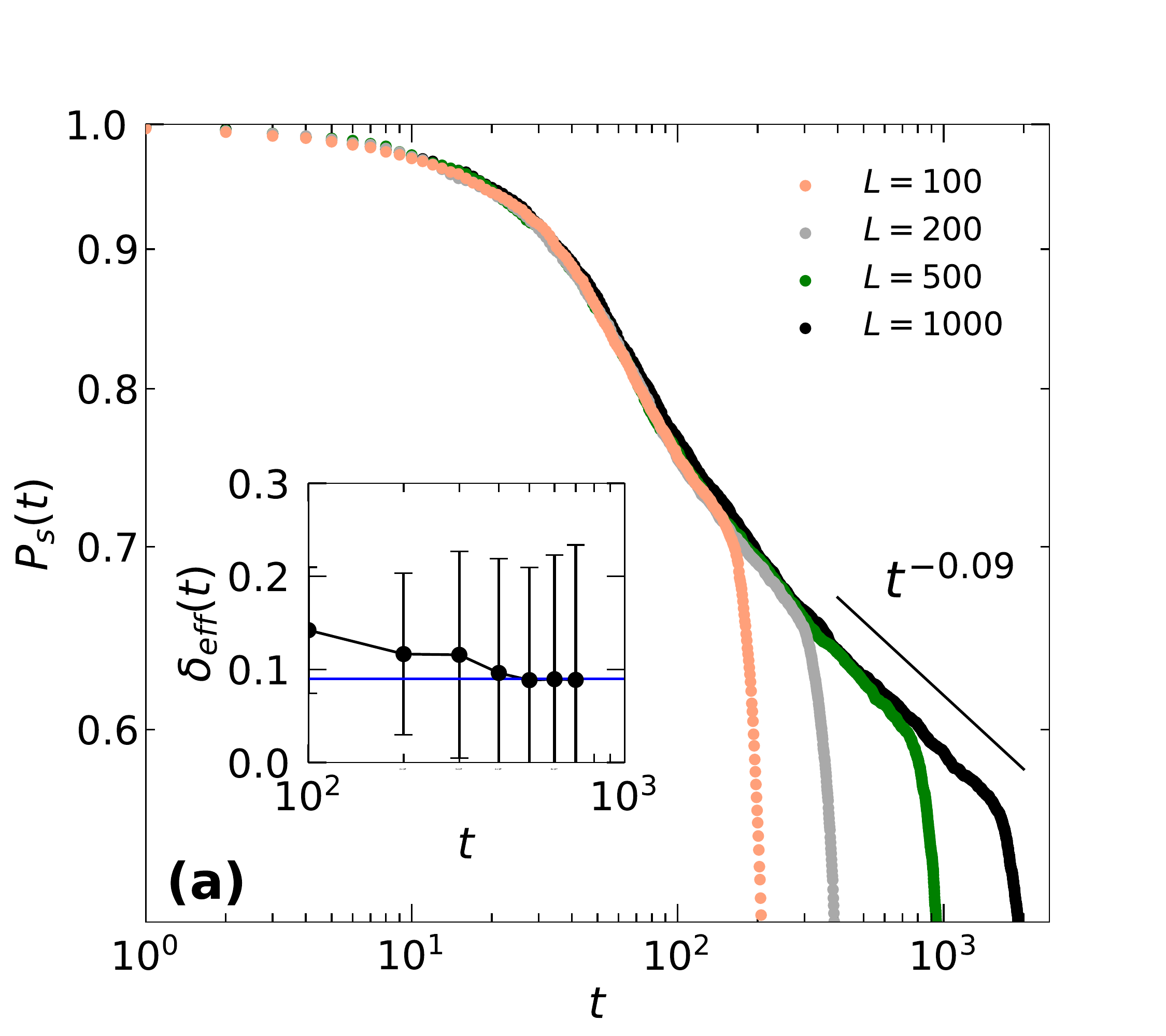}}
    \vfill
    \vspace{-0.5cm}
    \subfloat{\includegraphics[width=\columnwidth, trim={0cm 0 0.8cm 0.8cm},clip]{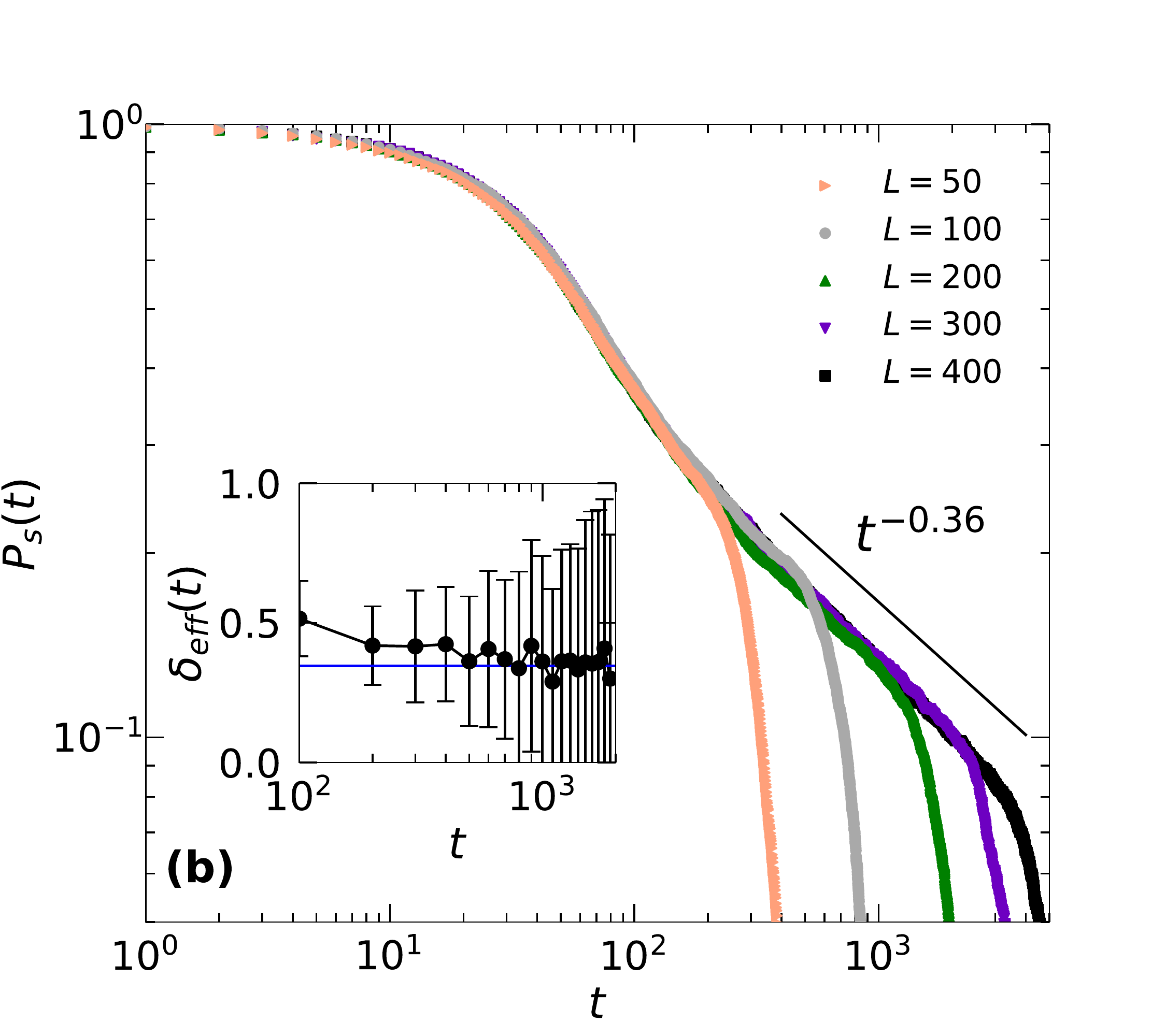}}
    \vspace{-0.3cm}
    \caption{Time evolution of the infection survival probability $P_s(t)$ at the critical point, measured at the site percolation threshold density $\rho^*$ for different system sizes of (a) square ($d = 2$) and (b) cubic ($d = 3$) lattices. 
    The insets show the corresponding  time-dependent effective exponents $\delta_{\rm eff}(t)$ for system sizes (a) $L = 1000$ and (b) $L = 400$. 
    The horizontal blue lines indicate the DyIP values of the critical survival exponent $\delta$. 
    The survival probability curves as well as the inset data in (a) were obtained from $5000$ independent realizations, while the data for the inset in (b) needed to be averaged over $20000$ independent simulation runs.} \label{fig:fig3}
\end{figure}
%
We observe that at densities close to $\rho = 1.0$, i.e., up to $\rho = 0.8$ in two and $\rho = 0.7$ in three dimensions, the simulation data look very similar to those depicted in Fig.~\ref{fig:fig1} for completely filled lattices. 
Measuring the (effective) critical exponents at these densities, we found their values to be identical (within the error bars) with the ``clean'' values. 
However, as the site percolation threshold density is approached, one initially discerns a markedly steeper decay of the survival probabilities for both square and cubic lattices. 
Right at the percolation threshold $\rho = \rho^*$, we see that this steeper power law region in the survival probability curve is quickly followed by an exponential cutoff which originates from the systems' finite sizes. 
As is apparent in Fig.~\ref{fig:fig3}, the steeper power law decay region might be mistaken as a signature of a different type of collective critical behavior; only after the simulation domain sizes are sufficiently increased does it become manifest that this apparently non-universal power law region in fact just represents a transient crossover regime. 
Both our best estimates of the asymptotic exponent $\delta = 0.09 \pm 0.005$ for $d = 2$ and $\delta = 0.36 \pm 0.02$ for $d = 3$ and the time-dependent effective exponent data $\delta_{\rm eff}(t)$ plotted in the insets of Fig.~\ref{fig:fig3} indicate that the exponents approach their universal DyIP values (from above) in the asymptotic time limit.

\begin{figure}[t!]\label{fig:fig4}
        \centering
    \subfloat[\label{fig:fig4a}]{\includegraphics[width=\columnwidth, trim={0.0cm 0 0.8cm 0.8cm},clip]{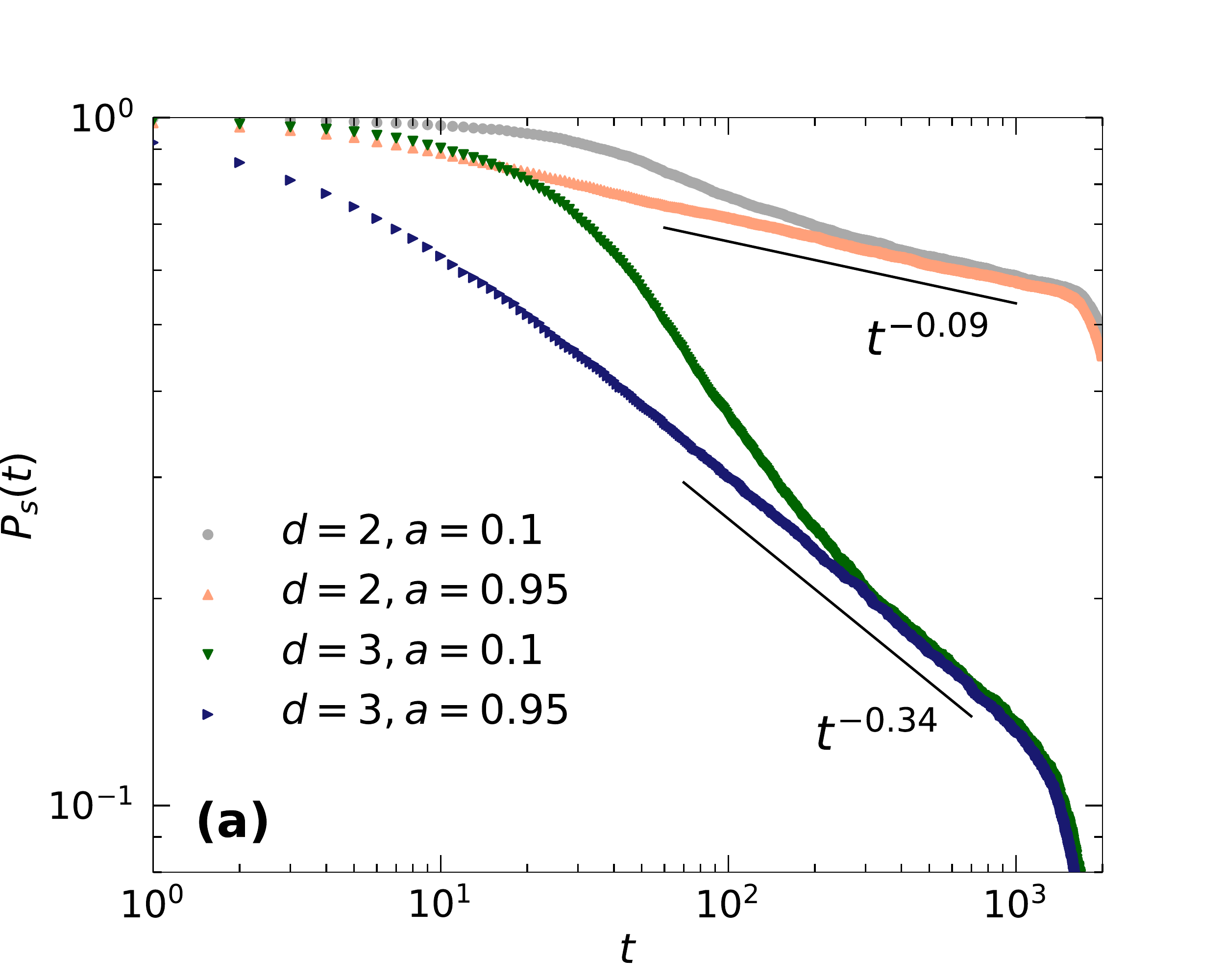}}
    \vfill
    \vspace{-0.5cm}
    \subfloat[\label{fig:fig4b}]{\includegraphics[width=\columnwidth, trim={0cm 0 0.8cm 0.8cm},clip]{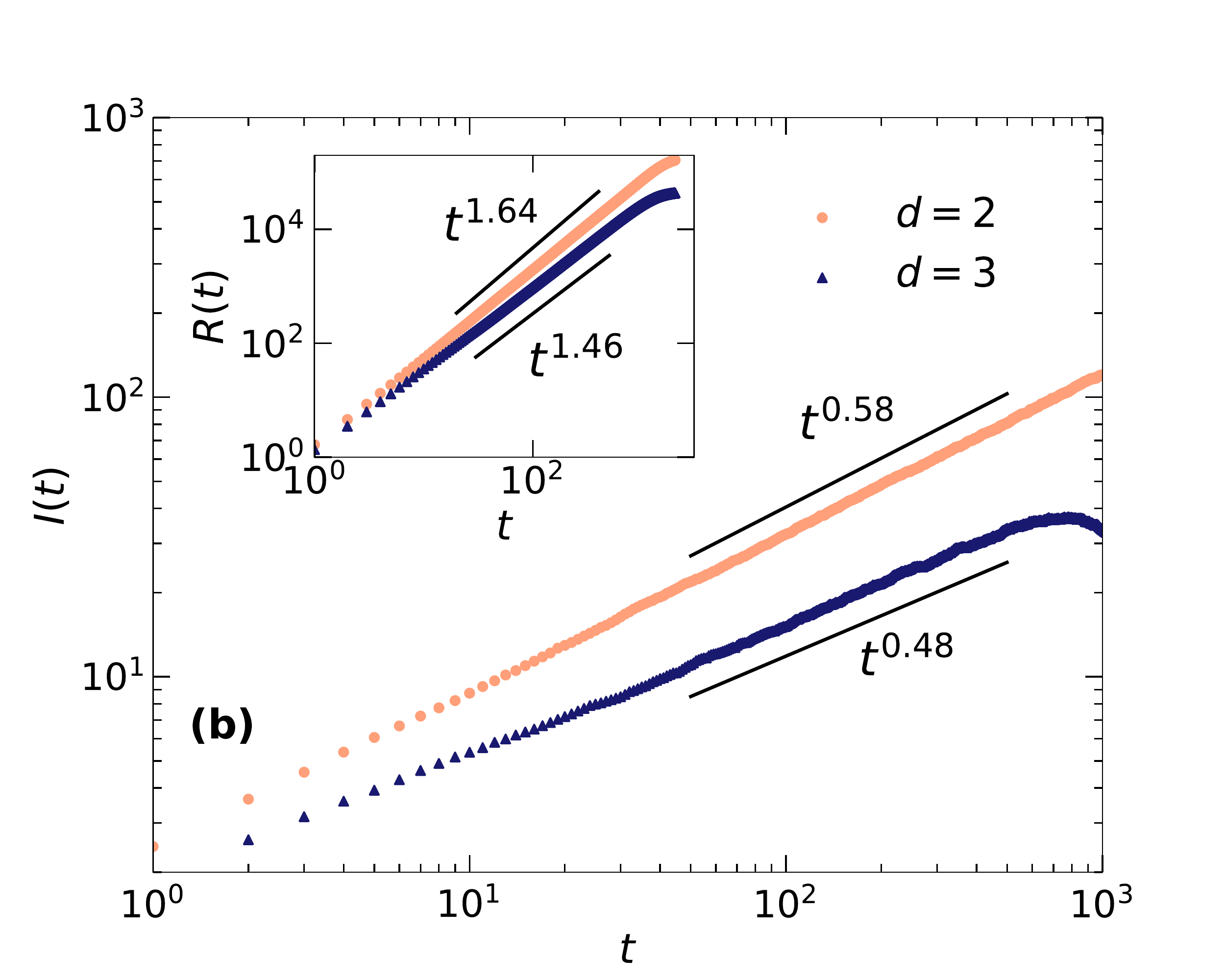}}
    \vspace{-0.3cm}
    \caption{(a) Time evolution of the infection survival probability $P_s(t)$ at the critical point, measured at the site percolation threshold density $\rho^*$ for different recovery rates. 
    (b) The time evolution of the total number of infected $I(t)$ and recovered individuals $R(t)$ (inset), obtained from the same data for $a=0.95$ recovery rate. 
    Our best estimates for the exponents are $\delta = 0.09 \pm 0.005$, $\theta = 0.58 \pm 0.01$, $\theta_\text{R} = 1.64 \pm 0.05$ for $d = 2$ and $\delta = 0.34 \pm 0.01$, $\theta = 0.48 \pm 0.01$, $\theta_\text{R} = 1.46 \pm 0.05$ for $d = 3$.
    The system sizes are $1000^2$ and $200^3$ for the two- and three-dimensional lattices, respectively, and all curves were obtained from averaging over $5000$ independent realizations.}
\end{figure}

We attribute the presence of these exceedingly long crossover regions observed at low densities $\rho$, i.e., large site dilutions, to the increase in the role of the recovery process as one approaches the percolation threshold density. 
Away from the percolation threshold density, variations in the recovery rate only horizontally shift the population curves and change the decay rate of the infectious curve, but they do not modify the critical exponents. 
However, once we approach the lattice percolation threshold, we have to increase the infectious rate dramatically to compensate for the drop in the overall connectivity of susceptible individuals. 
Because the infectious rate then becomes quite high, the incipient infection front quickly sweeps through the connected cluster of susceptible individuals. 
As a result, at early simulation times most of the individuals that have been left behind the infectious front will be still in the infectious state. 
While asymptotically the time when the cluster becomes completely recovered will be related to its size, for small clusters that were quickly traversed by the infection front, the extinction time will be set by the inverse recovery rate $1/a$.  
Therefore, we expect that as we increase the value of the recovery rate, while keeping it still less than the infection rate, the extinction delay that is introduced by fast propagation of infection should decrease and eventually vanish. 
We demonstrate this in Fig.~\ref{fig:fig4a} by comparing the survival probability curves that have been obtained for the recovery rates $a=0.1$ and $a=0.95$, respectively.
As is apparent in the graphs, for high recovery rates the crossover disappears entirely and we observe a power-law decay of the survival probability with the ``clean'' exponents. 
To further confirm that the non-equilibrium phase transition in the diluted SIR is governed by the standard DyIP exponents, we demonstrate in Fig.~\ref{fig:fig4b} that the total number of infected and the total number of recovered individuals both follow the expected ``clean'' growth scaling laws.

\begin{figure}[t!]\label{fig:fig5}
    \centering
        \subfloat[\label{fig:fig5a}]{\includegraphics[width=\columnwidth, trim={0.0cm 0 0.8cm 0.8cm},clip]{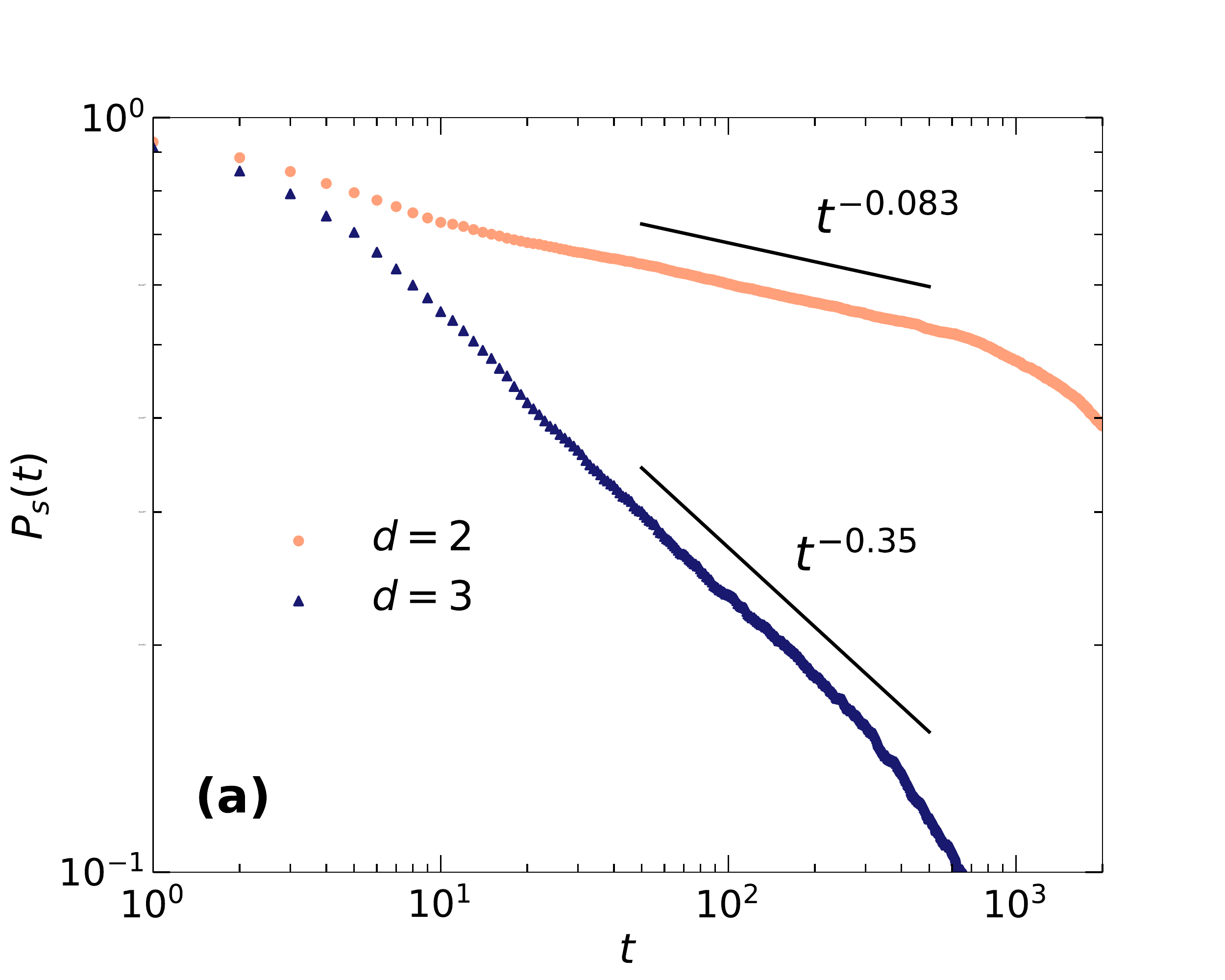}}
    \vfill
    \vspace{-0.5cm}
    \subfloat[\label{fig:fig5b}]{\includegraphics[width=\columnwidth, trim={0cm 0 0.8cm 0.8cm},clip]{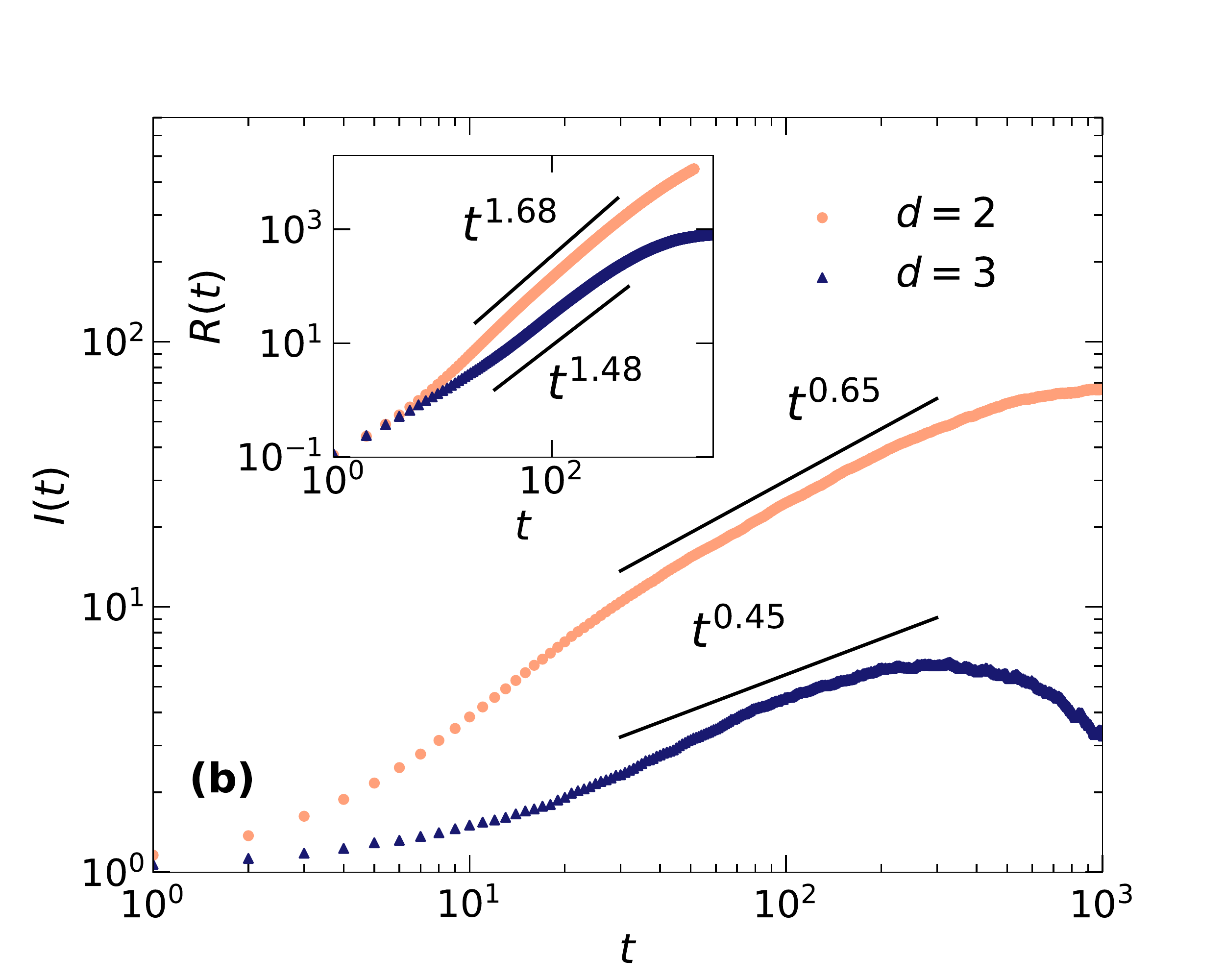}}
    \caption{(a) Time evolution of the infection survival probability $P_s(t)$ at the critical point for the SIR model on diluted lattices with enabled nearest-neighbor hopping, i.e., diffusive agent spreading. 
    (b) The time evolution of the total number of infected $I(t)$ and recovered individuals $R(t)$ (inset), obtained from the same data with enabled nearest-neighbor hopping. 
    The densities were chosen at the percolation thresholds $\rho^*$ in both two and three dimensions, see Fig.~\ref{fig:fig2}. 
    The critical values of the hopping, infectious, and recovery rates are $p_ c=1.0$, $r_c = 0.104$, $a_c = 0.1$ and $p_c=1.0$, $r_c = 0.093$, $a_c = 0.1$ for $d=2$ and $d=3$ respectively. 
    Our best estimates for the exponents are $\delta = 0.083 \pm 0.005$, $\theta = 0.65 \pm 0.1$, $\theta_\text{R} = 1.68 \pm 0.1$ for $d = 2$ and $\delta = 0.35 \pm 0.03$, $\theta = 0.45 \pm 0.1$, $\theta_\text{R} = 1.48 \pm 0.1$ for $d = 3$.
    The system sizes are $1000^2$ and $200^3$ for the two- and three-dimensional lattices, respectively, and the survival probability curves were obtained from $5000$ ($d = 2$) and $3000$ ($d = 3$) independent realizations.}
\end{figure}

%
Lastly, we investigate the more general and realistic situation when the agents on the lattice are no longer static, but allowed to hop on any empty nearest-neighbor sites with a non-zero hopping rate $p$. 
Choosing the hopping rate to be $p = 1.0$, we repeat the measurements of the same observables as before, for different density values. 
As seen in Fig.~\ref{fig:fig5}, we find that the critical infection rate values $r_c(\rho)$ are reduced significantly compared to the static scenario ($p = 0$). 
In contrast to what we observe in Fig.~\ref{fig:fig2}, for $p = 1.0$ the system will be in the spreading phase even for densities significantly lower than the site percolation threshold density $\rho^*$. 
Moreover, we notice in Fig.~\ref{fig:fig5} that the survival probability $P_s(t)$ does not exhibit the significant intermediate crossover region that is characteristic of the static case with immobile agents. 
In contrast to quenched disorder for which the impurity positions are held fixed, the agents' diffusive spreading dynamics quickly brings the system to an effectively uniformly diluted state. 
Therefore, from a coarse-grained point of view, the dynamics of this scenario of a site-diluted lattice with enabled nearest-neighbor hopping should resemble a normal isotropic percolation process at sufficiently large system size.
The only difference to a fully filled lattice setup is that the dynamics at the critical point $r_c$ for the case with enabled hopping is drastically slower due to the much lower critical infection rates. 
This renders the transient regimes quite long, obstructing our measurements of the ultimate critical scaling exponents, for which our best estimates are $\delta = 0.083 \pm 0.005$, $\theta = 0.65 \pm 0.1$, $\theta_\text{R} = 1.68 \pm 0.05$ for $d = 2$ and $\delta = 0.35 \pm 0.03$, $\theta = 0.45 \pm 0.1$, $\theta_\text{R} = 1.48 \pm 0.05$ for $d = 3$ (see Fig.~\ref{fig:fig5}). 

\section{\label{sec:level4} Conclusion}

In this work we have investigated the properties of the non-equilibrium active-to-absorbing state phase transition in the paradigmatic stochastic Susceptible-Infectious-Recovered epidemic model, where the SIR reactions are implemented on regular two- and three-dimensional square / cubic lattices with quenched site dilution disorder.
In the context of epidemic spreading (here with recovery), the effective elimination of susceptible sites could be caused by immunizations.
Even though the DyIP exponents that characterize the disorder-free transition satisfy the Harris criterion, we observe that for agent densities close to the percolation threshold the effective exponents seem to deviate from their ``clean'' values.

Only after adequately increasing its size can the system experience a crossover to the asymptotic scaling behavior, and the effective exponents assume their universal DyIP values after sufficiently long simulation times. We have shown that the crossover time can be reduced significantly upon increasing the recovery rate.

Consequently, due to the system's finite size there is a danger of mistaking transient impurity effects for apparent signatures of non-universal critical scaling. 
We note that this feature may be relevant for experiments, too, where imperfections are almost inevitable and for which finite-size effects may be especially prominent. 
We also provide an estimate for the characteristic crossover time when the system transitions to the asymptotic universal DyIP scaling behavior.

Finally, we have shown that when the SIR reactions are modeled on a diluted lattice with enabled nearest-neighbor agent hopping, the long crossover that we observe for the lattice with quenched disorder is no longer present, and the non-equilibrium active-to-absorbing phase transition is simply governed by the DyIP exponents.
In accord with Ref.~\cite{PhysRevE.103.062112}, our findings demonstrate the asymptotic irrelevance of quenched disorder on the critical properties of the dynamic isotropic percolation universality class for generalized epidemic spreading with recovery and immunity.
This is in stark contrast to the strong modifications induced by disorder on the related directed percolation universality class that describes simple epidemic processes (without recovery) near their threshold.

Here we were able to consistently measure only the critical exponent $\delta$ that characterizes the temporal power law of the survival probability curve. 
Other quantities such as the infection curve exhibit an early maximum that cuts off critical power laws, while we found the mean-square displacement data to be plagued by very noisy statistics. 
Therefore, to probe carefully the other critical exponents such as $\theta$ or $z$ in the presence of quenched disorder, one would need to run simulations for substantially greater system sizes and collect even more independent realizations; neither of which is currently feasible with the computational resources available to us.
It would also be interesting to further investigate the effects of Poisson-generated disconnected clusters which appear to induce the observed long crossover times, and to shed additional light on how the increase in lattice size facilitates the system to eventually exit this transient crossover regime. 

\begin{acknowledgments}
The authors would like to thank Shengfeng Deng, Geza \'Odor, and Michel Pleimling for insightful discussions.
This research was sponsored by the Army Research Office and was accomplished under Grant No. W911NF17-1-0156. 
The views and conclusions contained in this document are those of the authors and should not be interpreted as representing the official policies, either expressed or implied, of the Army Research Office or the U.S. Government. 
The U.S. Government is authorized to reproduce and distribute reprints for Government purposes notwithstanding any copyright notation herein.
\end{acknowledgments}

\bibliographystyle{apsrev4-1}
\bibliography{references}

\end{document}